\documentclass[twocolumn,aps,prl,showpacs,groupedaddress]{revtex4}

\usepackage{amsmath}
\usepackage{dcolumn}
\usepackage{epsfig}
\usepackage{graphicx}
\usepackage{latexsym}
\usepackage{pdfpages}

\usepackage{epstopdf}

\begin{document}

\title{Auto-Balanced Ramsey Spectroscopy}

\author{Christian Sanner, Nils Huntemann, Richard Lange, Christian Tamm, and Ekkehard Peik}
\affiliation{Physikalisch-Technische Bundesanstalt, Bundesallee 100,
38116 Braunschweig, Germany}

\begin{abstract}
We devise a perturbation-immune version of Ramsey's method of separated oscillatory fields. Spectroscopy of an atomic clock transition without compromising the clock's accuracy is accomplished by actively balancing the spectroscopic responses
from phase-congruent Ramsey probe cycles of unequal durations. Our simple and universal approach eliminates a wide variety of interrogation-induced line shifts often encountered in high precision spectroscopy, among them, in particular, light shifts, phase chirps, and transient Zeeman shifts. We experimentally demonstrate auto-balanced Ramsey spectroscopy on the light shift
prone $^{171}$Yb$^{+}$ electric octupole optical clock transition
and show that interrogation defects
are not turned into clock errors.
This opens up frequency
accuracy perspectives below the $10^{-18}$ level for the Yb$^{+}$ system and for other types of optical clocks.
\end{abstract}

\maketitle

A measurement of a physical observable in a quantum system profoundly affects the system's state.
%
%
Within the limits set by quantum mechanics \cite{Sakurai1994},
precision measurements strive
to minimize fluctuations and systematic errors due to measurement-induced perturbations.
In this regard atomic clocks are a prominent example: With their ever-improving stability
and accuracy \cite{Ludlow2015}
they have now reached a level of
performance that renders previously negligible clock errors
introduced with the spectroscopic interrogation highly relevant.

In this Letter we show that a balanced version of Ramsey's method of
separated oscillatory fields \cite{Ramsey1949,Ramsey1950} is well suited for measuring
unperturbed transition frequencies of ultra narrow atomic reference
transitions that suffer from significant
frequency shifts due to the
spectroscopic interrogation with the oscillatory drive pulses.
Relying on simple common-mode suppression arguments we devise an
auto-balancing scheme which unlike more specialized composite pulse
proposals \cite{Warren1983,Yudin2010,Hobson2016,Zanon2016,Yudin2016} and other modified Ramsey techniques \cite{Zanon2005,Hafiz2017,Vutha2015} provides universal immunity to
all kinds of interaction pulse aberrations and associated systematic
shifts including drive-induced AC Stark shifts (light shifts) \cite{Haeffner2003,Santra2005}, phase chirps and other phase and/or frequency deviations \cite{Falke2012}, transient Zeeman shifts \cite{Akatsuka2008}, and other pulse-synchronous shifts \cite{Rosenband2008}.

Using the example of the strongly light shift disturbed
$^{171}$Yb$^{+}$ electric octupole (E3)
optical clock transition at 467~nm \cite{Roberts2000}
we experimentally demonstrate the advantages of auto-balanced Ramsey
spectroscopy and show that no systematic clock errors are incurred
for arbitrarily detuned or otherwise defective drive pulses.
In addition we present in this context an experimental method
addressing spectroscopy-degrading issues related to the motional ion
heating \cite{Brownnutt2015} typically encountered in
ion traps.

Our work is motivated by the prospect of exploiting the full
potential of ultra narrow clock transitions without being limited by
the detrimental consequences of very weak oscillator strengths and
measurement-induced imperfections.
In particular for the Yb$^{+}$ E3 clock
this opens up the path to reproducible
long-term frequency ratio measurements as discussed in searches for
variations of fundamental constants \cite{Uzan2011,Godun2014,Huntemann2014var}, violations of Lorentz
invariance \cite{Dzuba2016}, and ultra-light scalar dark matter \cite{Stadnik2015}.

Ramsey spectroscopy
conceptually relies on the measurement of the
relative phase $\Delta \varphi$ accumulated in a superposition state
$|g\rangle + |e\rangle e^{-i \varphi(t)}$ over a free evolution time
$T$ (often called dark time or Ramsey time). ${\Delta \varphi = E_{e} T / \hbar}$ with $\hbar$ being the reduced
Planck constant, directly reflects the energy difference $E_{e}$ between
excited state $|e\rangle$ and ground state $|g\rangle$.
Ramsey spectroscopy compares this evolving atomic phase $\varphi(t)$ to the
phase evolution $\phi_{\mathrm{LO}} (t)$ of a local oscillator (LO),
e.g., a microwave source or a laser \cite{Footnote01}.
The fact that the spectroscopically relevant phase information is
acquired during an interaction-free period makes Ramsey's protocol
the natural choice when
aiming to undisturbingly extract the
transition frequency $\omega_{eg} = E_{e} / \hbar$. Nonetheless, the
deterministic preparation of the initial superposition state and the
subsequent phase readout which maps phase differences to directly
observable population differences require interactions with the
oscillatory drive pulses. Hence pulse defects (e.g., frequency
deviations) and other interrogation artifacts potentially affect the
outcome of the spectroscopic measurement leading to erroneous phase
values and corresponding clock errors.

To further
analyze this error propagation
we adopt the standard 2-by-2 matrix based
description of a coherently driven two-level system \cite{Sakurai1994,Yudin2010}. A drive field
oscillating with $\cos \phi_{\mathrm{LO}} (t)$ connects the two energy eigenstates $|g\rangle$ and
$|e\rangle$. Applying this field for a time $\tau = t_{1} - t_{0}$
and neglecting counter-rotating terms converts an initial state
$|i; t=t_{0}\rangle = g_{0} |g\rangle + e_{0} |e\rangle$ into a
final state $|f; t=t_{1}\rangle = g_{1} |g\rangle + e_{1} |e\rangle$
via $\begin{pmatrix}g_{1} \\ e_{1}\end{pmatrix} =$

\begin{equation}
\label{e:PROPA1} e^{i \delta' \tau/2} \:
\mathcal{V}[-\phi_{\mathrm{LO}} (t_{1})] \: \mathcal{U}[\delta', \tau,
\Omega_{0}] \: \mathcal{V}[\phi_{\mathrm{LO}} (t_{0})]
\begin{pmatrix}g_{0} \\ e_{0}\end{pmatrix}.
\end{equation}

In this expression the unitary matrices $\mathcal{U}$ and
$\mathcal{V}$ are defined as $\mathcal{U}[\delta', \tau, \Omega_{0}]
=$

\begin{equation}
\begin{pmatrix} \cos \frac{\Omega \tau}{2} - i \frac{\delta'}{\Omega} \sin \frac{\Omega \tau}{2} &
i \frac{\Omega_{0}}{\Omega} \sin \frac{\Omega \tau}{2} \\
i \frac{\Omega_{0}}{\Omega} \sin \frac{\Omega \tau}{2} &
\cos \frac{\Omega \tau}{2} + i \frac{\delta'}{\Omega} \sin \frac{\Omega \tau}{2}
\end{pmatrix}
\end{equation}

and $\mathcal{V}[\xi] = \begin{pmatrix} 1 & 0 \\0 & e^{i \xi}
\end{pmatrix}$ with Rabi frequency $\Omega_{0}$ and generalized Rabi
frequency $\Omega = \sqrt{\Omega_{0}^{2} + \delta'^{2}}$
characterizing the strength of the coupling between drive field and
atomic transition. $\delta' = \omega_{\mathrm{LOdrive}} -
\omega'_{eg}$ denotes the detuning of the oscillatory drive
frequency $\omega_{\mathrm{LOdrive}}$ from the instantaneous
atomic resonance frequency $\omega'_{eg}$. Primed variables identify
quantities
that due to interaction-induced shifts
possibly deviate from their unprimed
counterparts. For instance, in the case of the $^{171}$Yb$^{+}$
octupole transition $\omega_{eg}$ corresponds to the undisturbed
true clock transition frequency;
during the initialization and readout Ramsey pulses, however,
off-resonant coupling of the high intensity drive light to several
dipole transitions outside the two-level system leads to
a light-shifted instantaneous clock transition frequency
$\omega'_{eg}$ with an
unknown light shift $\Delta \omega_{eg} = \omega'_{eg} -
\omega_{eg}$ easily
exceeding $\Omega_{0}$.

In the following
we consider a standard optical clock operation scenario where an ultra-stable
laser is frequency-locked to the atomic reference transition via
successive
spectroscopic interrogations. The integrated outcome of the
individual excitation attempts
provides frequency feedback to the laser source.

Using the Ramsey scheme an ideal
interrogation sequence starting from the atomic ground state is then
comprised of three
phase-continuously connected segments:
First a drive pulse at
frequency $\omega_{\mathrm{LOdrive}} = \omega'_{eg}$ with
$\Omega_{0} \tau = \pi / 2$ (i.e., a $\pi/2$-pulse) initializes the
atomic superposition state. Then over a free evolution Ramsey time
$T$ the local oscillator's phase evolves with a rate
$\omega_{\mathrm{LO}} = \omega_{eg}$ that matches the undisturbed
atomic phase accumulation.
During this interval the LO
phase $\phi_{\mathrm{LO}} (t)$ is alternately
incremented or decremented by $\phi^{\pm} = \pm \pi / 2$ \cite{Ramsey1951}. Finally another $\pi/2$-pulse with
$\omega_{\mathrm{LOdrive}} = \omega'_{eg}$ is applied and
subsequently the excited state population $p^{\pm}$ is determined.
The $\phi^{\pm}$ modulation
offsets the normally
observed Ramsey fringe $p(\delta)$ (i.e., the final excited state
population as a function of Ramsey detuning $\delta =
\omega_{\mathrm{LO}} - \omega_{eg}$) by a quarter fringe period.
Accordingly, the differential excited state population $\tilde{p}
(\delta) = p^{+} (\delta) - p^{-} (\delta)$ can be interpreted as a
derivative-like frequency error signal with a zero crossing at the
Ramsey fringe center. Regulating $\omega_{\mathrm{LO}}$ such that
$\tilde{p} = 0$ closes the experimental feedback loop and
results in a locked local oscillator whose stability is ideally only
limited by the signal-to-noise ratio of the population measurements.

However, $\tilde{p}$ is not only a function of its explicit argument
$\delta$; as pointed out before, any aberration from the
ideal interrogation sequence, e.g., drive pulse defects (frequency-wise
quantified via $\delta' \neq 0$) or pulse-synchronous variations of the magnetic field,
also affect the outcome of the
population measurements and might
shift the error signal zero crossing point away from its undisturbed
position. Therefore an apparent $\tilde{p} (\delta) = 0$ does not
necessarily correspond to a true $\omega_{\mathrm{LO}} =
\omega_{eg}$ but can rather imply an actual clock error $\delta \neq
0$ whose magnitude will
then approximately scale inversely with $T$ \cite{Taichenachev2010}.

\begin{figure}[tbh!]
\begin{center}
\includegraphics[width=3.4in]{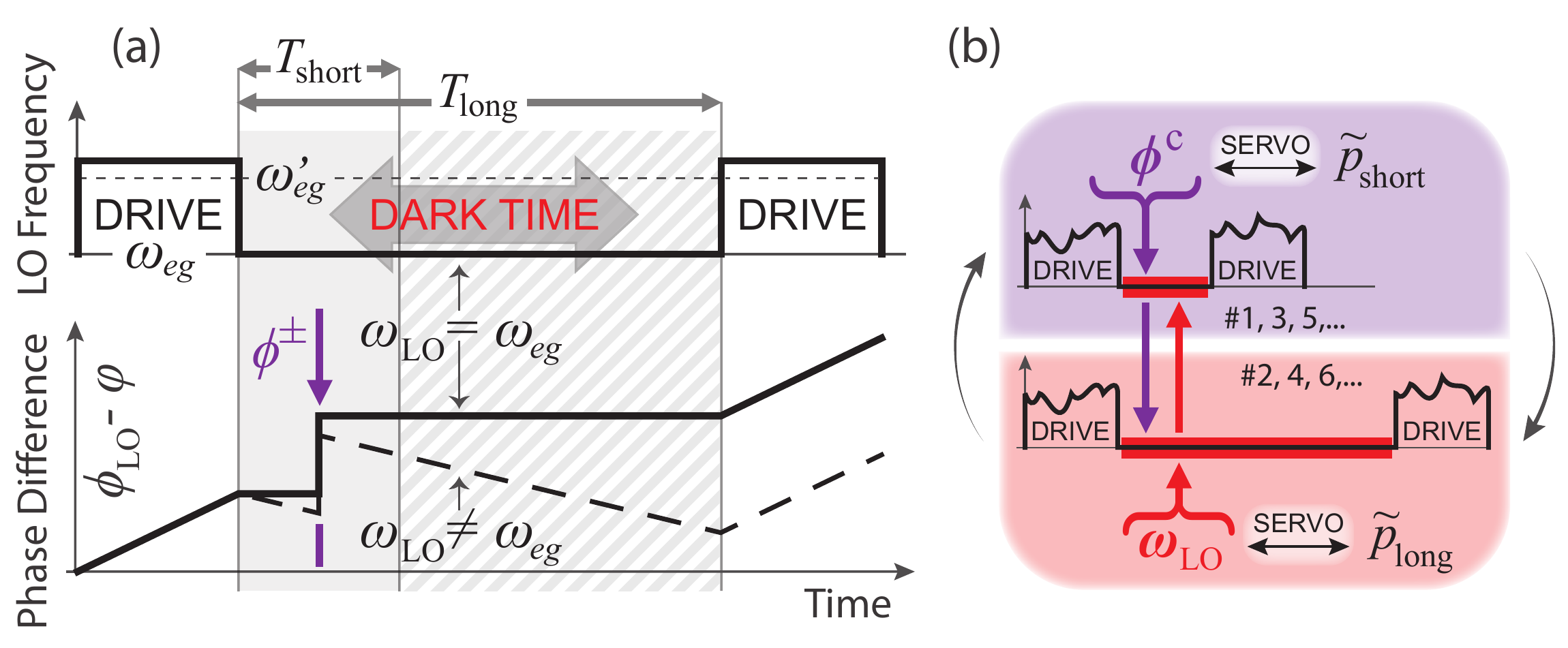}
\caption[]{Ramsey interrogation with shift-inducing drive pulses. Unlike in the ideal case where $\omega_{\mathrm{LOdrive}} = \omega'_{eg}$, the drive pulse frequency in (a) is assumed to be slightly higher than the light-shifted transition frequency $\omega'_{eg}$. Therefore, the LO phase evolves initially faster than the atomic superposition phase $\varphi(t)$. Over the following dark Ramsey time interval of variable duration $T$ the acquired phase difference between LO field and atomic oscillator will remain unchanged (aside from the $\phi^{\pm}$ modulation) if the LO phase advances with the unperturbed transition frequency $\omega_{\mathrm{LO}} = \omega_{eg}$ during $T$. In this case the phase-to-population conversion performed by the second drive pulse will produce identical outcomes after short and long Ramsey times. (b) Relying on this $T$-invariant response an auto-balancing control scheme makes it possible to extract the unperturbed atomic transition frequency even with arbitrarily distorted drive pulses. The servo-controlled phase correction $\phi^{\mathrm{c}}$ balances the short sequence and immunizes the interconnected LO-steering long sequence against drive-induced shifts.\label{f:pulsesandloops}}
\end{center}
\end{figure}

To correctly reproduce $\omega_{eg}$ it is essential to
distinguish deviations of $\tilde{p}$ due to $\omega_{\mathrm{LO}}
\neq \omega_{eg}$ from misleading deviations
caused by a flawed interrogation process.
This discrimination is
accomplished by exploiting the above mentioned dependance of
$\delta$ and thus $\tilde{p}$ on $T$: As shown in Figure
\ref{f:pulsesandloops}(a), only for $\omega_{\mathrm{LO}} =
\omega_{eg}$, i.e., only if LO phase and atomic phase evolve
at identical pace, one can expect to find the same excited state
population when comparing the outcome of two "isomorphic" interrogation sequences
which only differ in their dark times $T_{\mathrm{short}}$ and
$T_{\mathrm{long}}$ (see also reference \cite{Morgenweg2014}).

Based on the simple insight that interrogation-induced $\tilde{p}$ deviations are common mode for isomorphic Ramsey sequences it is now straightforward to combine $T_{\mathrm{short}}$ and $T_{\mathrm{long}}$ Ramsey
interrogations yielding $\tilde{p}_{\mathrm{short}}$ and
$\tilde{p}_{\mathrm{long}}$, respectively, into a defect-immune
spectroscopy scheme: By using the differential
population $\tilde{p}_{\mathrm{bal}} = \tilde{p}_{\mathrm{long}} -
\tilde{p}_{\mathrm{short}}$ as an error signal for
$\omega_{\mathrm{LO}}$ one obtains a passively balanced frequency
feedback with $\tilde{p}_{\mathrm{bal}} = 0$ exclusively for
$\omega_{\mathrm{LO}} = \omega_{eg}$.

\begin{figure}[tbh!]
\begin{center}
\includegraphics[width=3.4in]{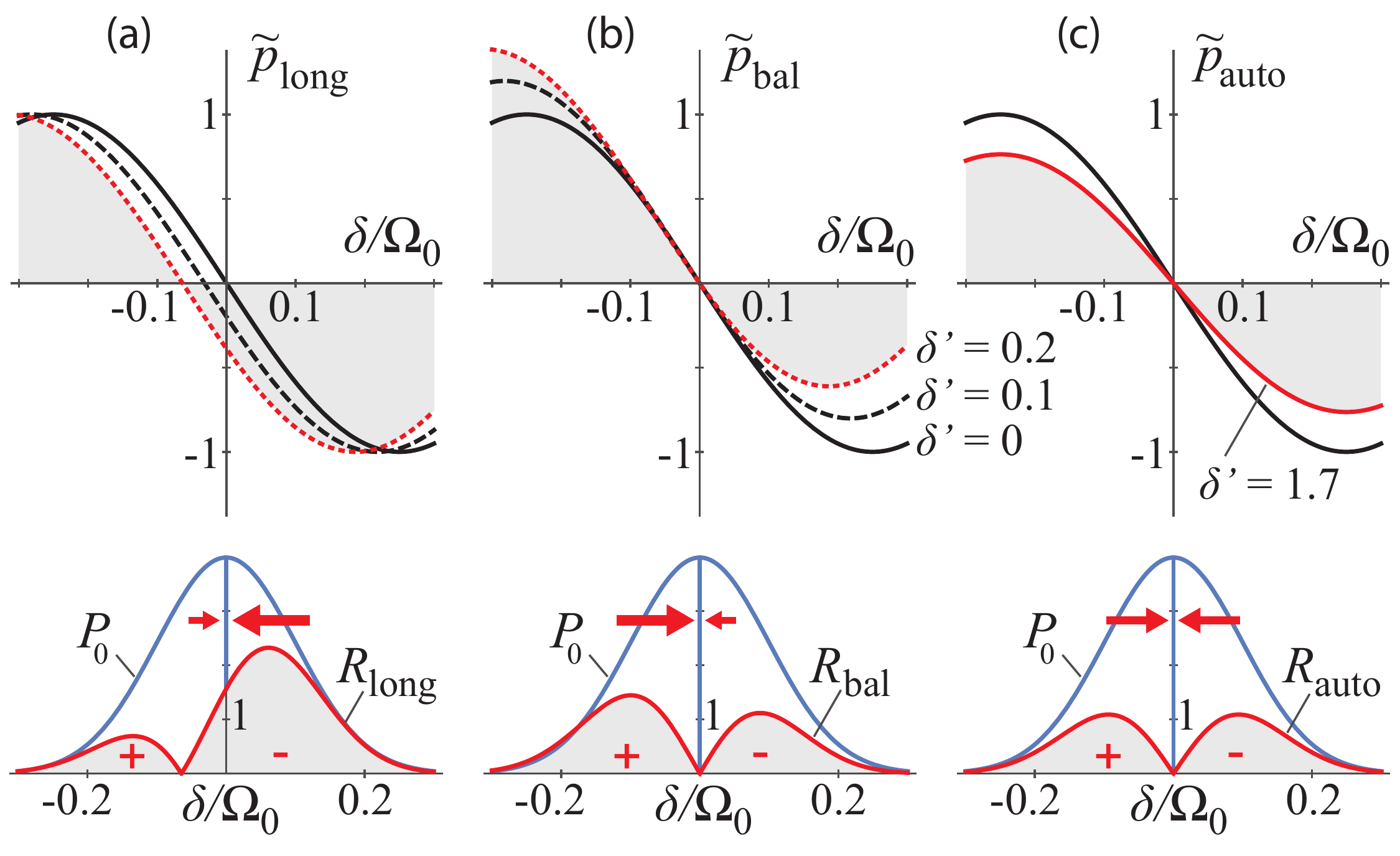}
\caption[]{Simulated Ramsey lock fringes $\tilde{p}(\delta)$ (top row) and corresponding rectified feedback responses $R(\delta) = |\tilde{p} P_{0}|$ for a resonant Gaussian sampling frequency distribution $P_{0}(\delta)$ (bottom row). Solid, dashed and dotted lines represent fringes with $\delta' / \Omega_{0}= 0$, $\delta' / \Omega_{0}= 0.1$ and $\delta' / \Omega_{0}= 0.2$, respectively. All curves are calculated assuming $\Omega_{0} T_{\mathrm{long}} = 2 \pi$. (a) Fringes obtained via the standard Ramsey protocol are horizontally shifted by approximately $\delta = -2 \delta' / (\Omega_{0} T)$ for small drive detunings $\delta'$. Accordingly, an engaged feedback loop will pull the frequency distribution toward lower frequencies. (b) Passive balancing displaces the fringe laterally so that its zero crossing is kept at the origin. However, for $\delta' \neq 0$ any finite-width sampling distribution will experience a skewed response that pushes the center frequency away from the $\delta = 0$ lock point. (c) Auto-balanced generation of the frequency discriminant ensures antisymmetric frequency feedback for any drive detuning. At large detunings $\delta' > \Omega_{0}$ the fringe looses contrast but maintains all symmetry properties.\label{f:fringes}}
\end{center}
\end{figure}

This approach, however, comes with a major
drawback, that is to a lesser extent also
encountered in recently proposed coherent composite pulse schemes
\cite{Hobson2016,Zanon2016}: For $\tilde{p}_{\mathrm{short}} \neq 0$
the frequency discriminant $\tilde{p}_{\mathrm{bal}}
(\delta)$ is no longer an odd function around $\delta = 0$, which
leads to skewed sampling distributions and corresponding clock
errors as explained in Figure \ref{f:fringes}. To avoid this issue we
implement an \textit{active} balancing process with two interconnected
control loops as schematically displayed in Figure \ref{f:pulsesandloops}(b). The first feedback loop, acting on the
short Ramsey sequence,
ensures $\tilde{p}_{\mathrm{short}} = 0$ by injecting together with
$\phi^{\pm}$ an additional
phase correction $\phi^{\mathrm{c}}$. Of course, this requires
$\delta' < \Omega_{0}$
for sufficient fringe contrast.
Depending on the dominant
source of error one could also use $\omega_{\mathrm{LOdrive}}$ as
the control variable \cite{Footnote02}.
During $T_{\mathrm{short}}$ the local oscillator evolves with
$\omega_{\mathrm{LO}}$ as determined via the second feedback loop
which
controls the long Ramsey sequence by steering $\omega_{\mathrm{LO}}$
so that $\tilde{p}_{\mathrm{long}} = 0$. Vice versa,
$\phi^{\mathrm{c}}$ (or $\omega_{\mathrm{LOdrive}}$) as obtained
from the short sequence is identically applied in the long Ramsey
sequence. In this way a constantly updated common-mode correction
auto-balances the long interrogation whose outcome is then
denoted by $\tilde{p}_{\mathrm{auto}}$. As shown in Figure \ref{f:fringes}(c), auto-balancing
prevents skewed sampling responses and leads to an always
antisymmetric lock fringe $\tilde{p}_{\mathrm{auto}} (\delta)$ which
ensures that
deviations $\tilde{p}_{\mathrm{auto}} \neq 0$
are solely due to $\omega_{\mathrm{LO}} \neq \omega_{eg}$.

We now report on an application of auto-balanced Ramsey
spectroscopy with an ytterbium single-ion clock where
the spectroscopic interrogation
causes a particularly large perturbation of the atomic reference.
So far, optical clocks operating on the 467 nm E3 transition
in $^{171}$Yb$^{+}$ have employed extrapolation methods \cite{Godun2014,Huntemann2012} or
Hyper-Ramsey composite pulse approaches \cite{Huntemann2012hrs} to address the
drive light induced AC Stark shift.
Our experiments were carried out with the single $^{171}$Yb$^{+}$
ion confined in an endcap-type radio frequency Paul trap \cite{Stein2010}
whose drive amplitude was adjusted to provide radial secular motion
frequencies of $\omega_{r} = 2 \pi \times 1$ MHz and axial
confinement with $\omega_{a} = 2 \pi \times 2$ MHz.
Laser cooling is performed on the $^{2}S_{1/2}$ $\leftrightarrow$
$^{2}P_{1/2}$ cycling transition at 370 nm with repump lasers at 935
nm and 760 nm preventing population trapping in the metastable
$^{2}D_{3/2}$ and $^{2}F_{7/2}$ manifolds \cite{Huntemann2016}.

While not actively cooled the trapped ion linearly gains motional
energy \cite{Brownnutt2015} along each dimension, in our setup at a rate of 300 $\hbar \omega_{r}$ per
second.
This large heating rate
reduces the effective Rabi frequencies for Ramsey pulses applied
after long dark times.
To compensate for this effect one cannot simply change the laser light
intensity because $\omega'_{eg}$ would change accordingly.
Intensity-neutral $\Omega_{0}$
equalization is instead achieved
by changing the spectral composition of the drive light
with an electro-optic modulator that
redistributes light from the carrier into sidebands
spaced in our case $\omega_{\mathrm{EOM}} = 2 \pi \times 2$ GHz apart. The value for $\omega_{\mathrm{EOM}}$ can be chosen
over a wide range as long as the AC Stark shift's spectral variation is insignificant and the added sidebands do not interfere with the resonant population transfer.
Activating the modulator during the second Ramsey pulse of the short
sequence and properly adjusting the modulation depth equalizes the short and long sequences and recovers their
isomorphism. A modulation index of 0.6 rad was used to match the thermally averaged 10 \% reduction of $\Omega_{0}$ due to heating in the long sequence.

Except for
heating compensation and
interrogation specifics the optical clock is operated following the
procedure outlined
in reference \cite{Huntemann2012}. To facilitate measurements with well-controlled
detunings $\delta'$ the LO laser (pre-stabilized
to a high finesse optical cavity) is
referenced to an independent Yb$^{+}$ E3 clock setup.

\begin{figure}[tbh!]
\begin{center}
\includegraphics[width=3.4in]{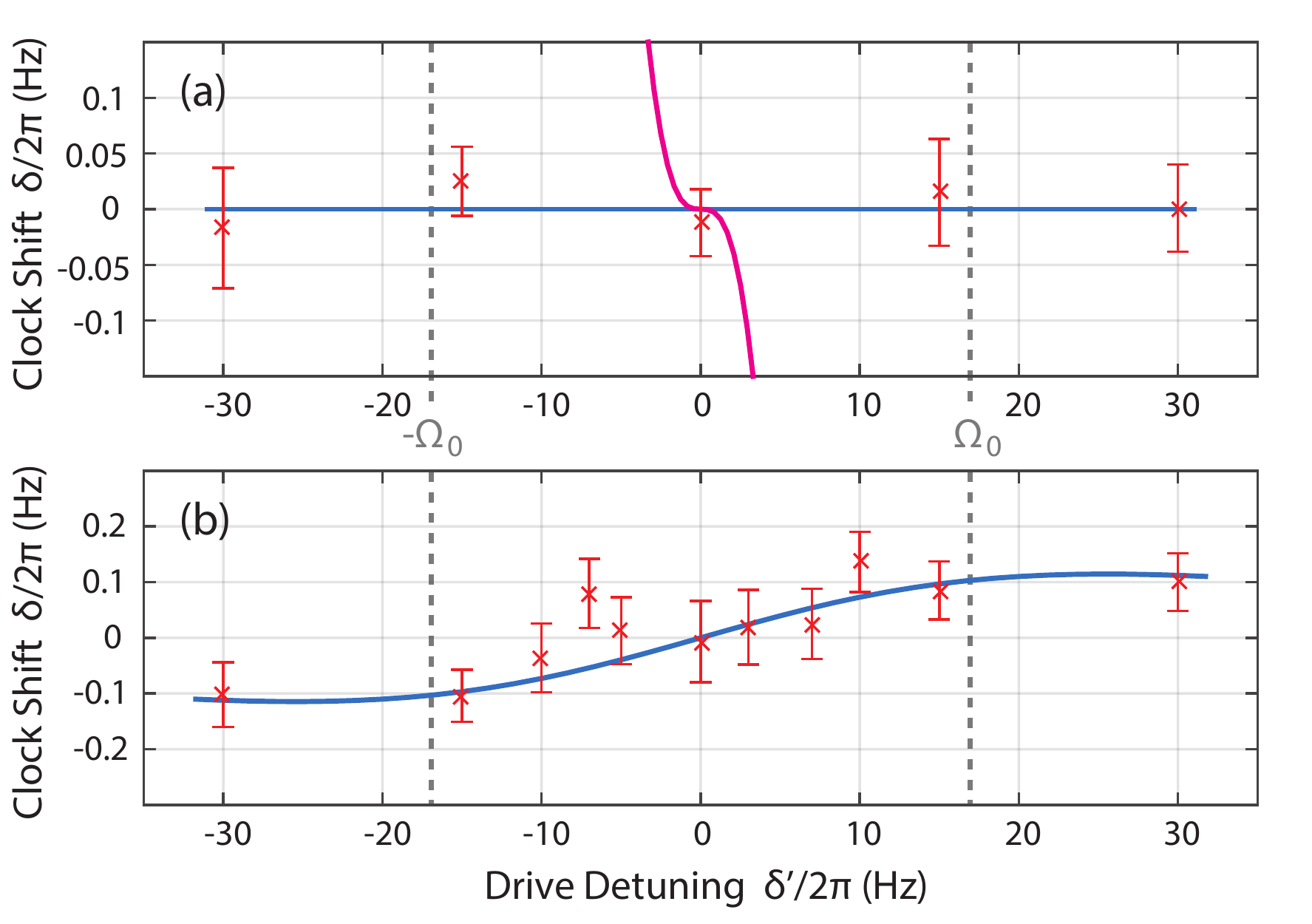}
\caption[]{Clock shifts on the $^{171}$Yb$^{+}$ E3 transition frequency as obtained through auto-balanced Ramsey spectroscopy when operated with intentionally detuned drive pulses. (a) For isomorphic short and long interrogation sequences the auto-balanced clock is fully immune against drive frequency deviations and the measured data points ($1\sigma$ error bars shown) line up on the blue zero-crosstalk axis. The purple curve illustrates the $\delta'$-to-$\delta$ error propagation expected with the original Hyper-Ramsey protocol. (b) Even without heating compensation the coupling between clock shift and drive detuning is strongly suppressed. Within statistical uncertainty the measured clock deviations reproduce the numerically simulated dependance (blue curve).\label{f:transfercurves}}
\end{center}
\end{figure}

Figure \ref{f:transfercurves}(a) shows the results of auto-balanced clock runs
with $\Omega_{0} = 2 \pi \times 17$ Hz corresponding to a $\pi /
2$-pulse duration of 15 ms, $T_{\mathrm{short}} = 6$ ms,
and $T_{\mathrm{long}} = 60$ ms.
Using about 5 mW of drive laser light focused
in a 50 $\mu$m diameter spot causes
a large light shift of $\omega'_{eg} - \omega_{eg} \approx 2 \pi
\times 660$ Hz.
Within the statistical uncertainty no clock error is
observed
for
drive detunings $\delta'$ of up to $2 \Omega_{0}$. Beyond
this detuning the response curve stays flat
but the data points'
statistical uncertainties eventually increase
due to the reduced fringe amplitude.
In contrast, operating a Hyper-Ramsey interrogation (REFS) at
$\delta' > \Omega_{0}$ yields
large clock errors that scale linearly with $\delta'$; only for
$\delta' \ll \Omega_{0}$ the error propagation is suppressed to
first order.

Without
heating compensation
one finds a residual dependance of $\delta$ on $\delta'$ as
displayed in Figure \ref{f:transfercurves}(b). This dependance is rather
weak and does not introduce a clock error as long as
$\delta'$ is on average zero; yet it confirms that heating violates the isomorphism of short and long Ramsey sequences.
Numerical simulations based on the thermally averaged outcome of
the multiplication of propagation matrices as introduced
in equation (\ref{e:PROPA1}) reproduce
the experimental results.

\begin{figure}[tbh!]
\begin{center}
\includegraphics[width=3.4in]{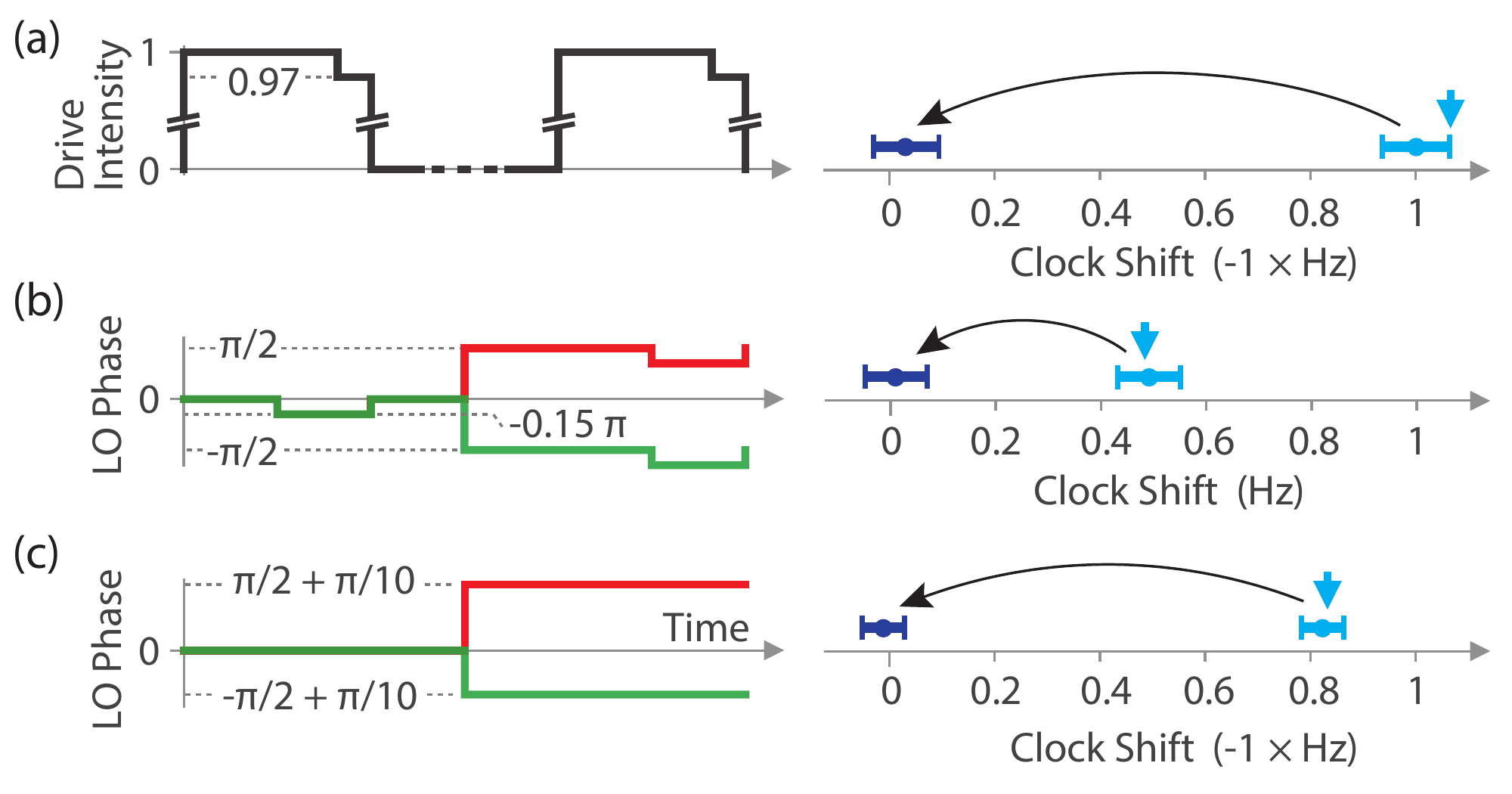}
\caption[]{Defective drive sequences and corresponding clock shifts measured via standard Ramsey spectroscopy (light blue data points with vertical arrows indicating the theoretically expected shift values) and via auto-balanced Ramsey spectroscopy (dark blue data points). The applied intensity defect (a), phase excursion (b), and phase lag (c) lead to large clock offsets that are eliminated in the auto-balanced acquisition mode. Both phase plots display $\phi^{-}$ (green) and $\phi^{+}$ (red) traces simultaneously.\label{f:pulsedefects}}
\end{center}
\end{figure}

To some extent, auto-balanced Ramsey spectroscopy and
Hyper-Ramsey spectroscopy can be interpreted as the incoherent and
coherent versions of the same underlying concept of common-mode
suppression, however, only the auto-balanced approach can provide
universal immunity to arbitrary interrogation defects. A
supplementary discussion found in \cite{Supplement} further explains the
analogy. In order to
illustrate this universality of the auto-balancing approach we
measured the Yb$^{+}$ clock's response to various intentionally
introduced interrogation defects. Figure \ref{f:pulsedefects}
displays three defect scenarios together with their resulting clock
errors. In the first scenario the $\pi / 2$-pulses are delivered
with 97 percent of the nominal intensity for the last 3 ms of
their 15 ms on-time.
Considering a total light shift of 660 Hz this intensity defect is
equivalent to a temporary $\omega_{\mathrm{LOdrive}}$ deviation of
more than $\Omega_{0}$ and gives rise to a clock shift of about 1
Hz when using an unbalanced Ramsey sequence.
Auto-balancing the acquisition
recovers the undisturbed clock frequency to within the statistical
uncertainty. Similarly,
defect immunity is
verified for drive pulses suffering from engineered phase defects
with 0.15 $\pi$ phase excursions during the second half of each
atom-light interaction.
Finally, a third scenario assumes a phase lag $\theta = \pi / 10$
after the first $\pi / 2$-pulse, i.e.,
one effectively
uses $\phi^{\pm} = \pm \pi / 2 + \theta$ instead of employing a
symmetric phase modulation. In this case the injected servo-controlled
phase correction $\phi^{\mathrm{c}}$ one-to-one
compensates the encountered phase lag.

While all these pulse defects are exaggerated for demonstration
purposes they represent
a wide variety of less pronounced and often unnoted
parasitic interrogation side effects. For instance, many clocks \cite{Berkeland2002} incorporate some form of magnetic field switching for the
atomic ground state preparation and certain prohibitively weak
clock transitions \cite{Barber2006} require magnetic field-induced state admixtures \cite{Taichenachev2006} to enable direct optical Ramsey
excitation.
Due to transient switching effects
a presumably dark Ramsey time might get contaminated.
Phase excursions (phase chirps) triggered by laser beam shutters or acousto-optic modulators are
another prominent source of error. By choosing a proper
$T_{\mathrm{short}}$ that still covers the transient behavior one
can preventively address all such issues.

Being focused on accuracy improvements we have
not considered
the implications of auto-balanced Ramsey spectroscopy for the
stability of atomic clocks. A detailed stability discussion
can be found in the supplemental material \cite{Supplement}.

In conclusion, we have introduced a conceptually simple and powerful spectroscopic technique to remove accuracy constraints imposed by interrogation-induced clock shifts. We have validated this technique using an optical ion clock and applied it to several highly disturbed Ramsey pulse sequences. This technique
is directly applicable to other atomic clocks and could be extended in the future to also address
quadrupole-mediated frequency errors \cite{Itano2000} and related
non-scalar shifts.

We thank Valery I. Yudin for valuable discussions, Burghard Lipphardt for experimental assistance, and Uwe Sterr for critical reading of the manuscript. This research has received funding from the EMPIR program co-financed by the Participating States and from the European Union's Horizon 2020 research and innovation program.


\pagebreak
\onecolumngrid
\begin{center}
\textbf{\large Auto-Balanced Ramsey Spectroscopy: Supplementary Information}
\vspace{0.8cm}
\end{center}

\twocolumngrid

\setcounter{equation}{0}
\setcounter{figure}{0}
\setcounter{table}{0}
\setcounter{page}{1}

\renewcommand{\theequation}{S\arabic{equation}}
\renewcommand{\thefigure}{S\arabic{figure}}
\renewcommand{\bibnumfmt}[1]{[S#1]}
\renewcommand{\citenumfont}[1]{S#1}

In this supplement we present a conceptual comparison of
coherent and incoherent pulse defect
immunization approaches.
Coherent approaches try to
implement frequency discrimination and shift suppression within a single phase-coherent interrogation cycle, while incoherent concepts rely on the combined spectroscopic outcome of separate interrogations.
We review the prototypical coherent Hyper-Ramsey interrogation
and contrast it with the incoherent auto-balanced Ramsey
interrogation protocol. Furthermore we discuss
implications of the different immunization schemes for the frequency
stability of the atomic clock. Variables are adopted as defined
in the main article.

\section{Coherent Defect-Immunization}

A coherent pulse defect immunization scheme attempts to
replace the standard $\pi/2$ Ramsey drive pulses with suitable composite pulse (CP) sequences so that the differential excited state population $\tilde{p}(\delta = 0)$ is only minimally affected when pulse parameters (e.g., $\delta'$) deviate from their
intended values. A wide variety of such CP sequences have been
employed
in NMR spectroscopy \cite{Levitt1986} or for high fidelity
ion addressing \cite{Ivanov2011}.
Beyond their specific
response plateaus the
pulse sequences have to meet additional symmetry criteria and
should not compromise spectroscopic sensitivity in order to serve as proper clock drive pulses.
In principle, a CP sequence (e.g., as found in \cite{Torosov2011}) can be turned into a defect-immune
interrogation scheme by inserting a dark Ramsey time $T$ somewhere
in the sequence. Two aspects
are emphasized: First, for $\delta = 0$ the addition of the Ramsey interval will not affect the measured outcome of the pulse sequence.
Second, the measured excited state population is the only parameter relevant for controlling $\omega_{\mathrm{LO}}$.

Accordingly, one can characterize the performance of a CP sequence employed for Ramsey superposition state preparation and readout
by choosing
$T=0$ and inspecting
how the final excited state populations $p^{+}$ and $p^{-}$ change as a function of
certain pulse
parameter deviations. We illustrate
this procedure by applying it to the Hyper-Ramsey CP protocol as introduced in \cite{Yudin2010}.

\section{The Hyper-Ramsey protocol}

\begin{figure}[tbh!]
\begin{center}
\includegraphics[width=3.4in]{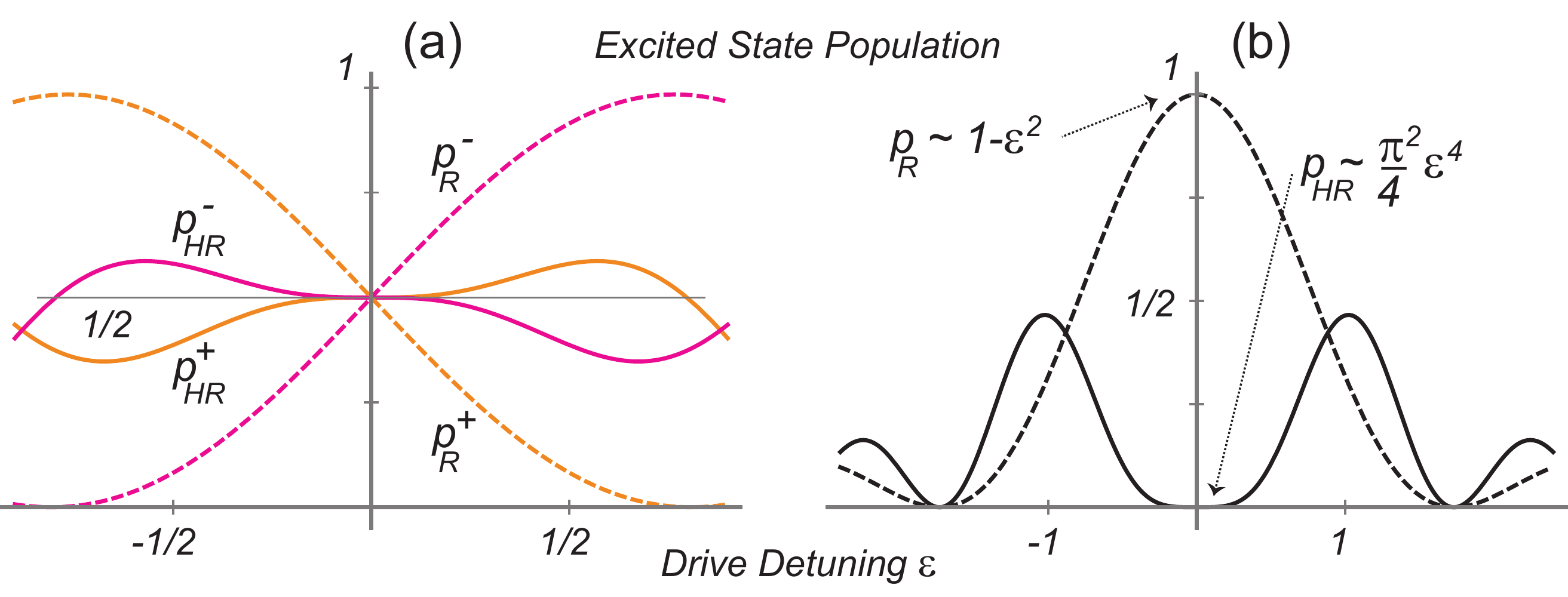}
\caption[]{Calculated two-level system excited state populations plotted versus drive detuning $\varepsilon = \delta' / \Omega_{0}$ in the absence of relaxation. (a) Around $\varepsilon = 0$ a standard Ramsey-type sequence with a $\phi^{+} = \pi/2$ ($\phi^{-} = -\pi/2$) phase hop between the two $\pi/2$-pulses exhibits a linear dependance of the population $p_{\mathrm{R}}^{+}$ ($p_{\mathrm{R}}^{-}$) on $\varepsilon$. For a Hyper-Ramsey-type sequence the population $p_{\mathrm{HR}}^{+}$ ($p_{\mathrm{HR}}^{-}$) is to first order unaffected by $\varepsilon$-deviations. (b) Without phase hops the corresponding populations $p_{\mathrm{R}}$ and $p_{\mathrm{HR}}$ reflect the spectroscopic outcome of simple $\pi$ and $2\pi$ pulses and scale quadratically and quartically with $\varepsilon$, respectively.\label{f:pplots}}
\end{center}
\end{figure}

In its generic form the Hyper-Ramsey scheme leaves the first pulse ($\pi/2$ preparation pulse) unchanged and triples the duration of the second pulse (readout pulse) so that $\Omega_{0} \tau = 3 \pi / 2$.
With $T=0$ one finds dependances of $p^{\pm}$ on the normalized drive pulse detuning $\varepsilon = \delta' / \Omega_{0}$ as displayed in Figure \ref{f:pplots}(a). Around $\varepsilon = 0$ the probabilities $p^{\pm}$ stay close to $1/2$
clearly indicating
the enhanced immunity of the CP sequence compared to a standard interrogation scenario with two $\pi/2$ pulses. This improvement can be understood intuitively by considering the two sequences with omitted
$\phi^{\pm}$ modulation. In this case the driven system performs simple Rabi oscillations given by
\begin{equation}
\label{e:pRabi} p(\tau) =\frac{1}{1 + \varepsilon^{2}}
\sin^{2} [\frac{\Omega_{0} \tau}{2} \sqrt{1 + \varepsilon^2}].
\end{equation}
The excited state population $p$ is probed after a total interaction
time $\tau_{\mathrm{HR}} = 2 \pi / \Omega_{0}$ (Hyper-Ramsey tripled
readout pulse duration) or $\tau_{\mathrm{R}} = \pi / \Omega_{0}$
(Ramsey $\pi/2$ readout pulse).

A deviation $\delta' \neq 0$ affects $p$ by means of an amplitude
"tilt" factor $1 / (1 + \varepsilon^{2})$ and an angular
"speed" factor $\sqrt{1 + \varepsilon^2}$. Figure
\ref{f:pplots}(b) illustrates that
the resulting $p$ deviation is of fourth order
in $\varepsilon$ for $\tau_{\mathrm{HR}}$-long pulses
and of second order
for $\tau_{\mathrm{R}}$-long pulses. This superior robustness of $2
\pi$ (and integer multiples of $2 \pi$) long pulse sequences along
with an unchanged spectroscopic sensitivity
is preserved when adding the $\phi^{\pm}$ modulation.

\begin{figure}[tbh!]
\begin{center}
\includegraphics[width=3.4in]{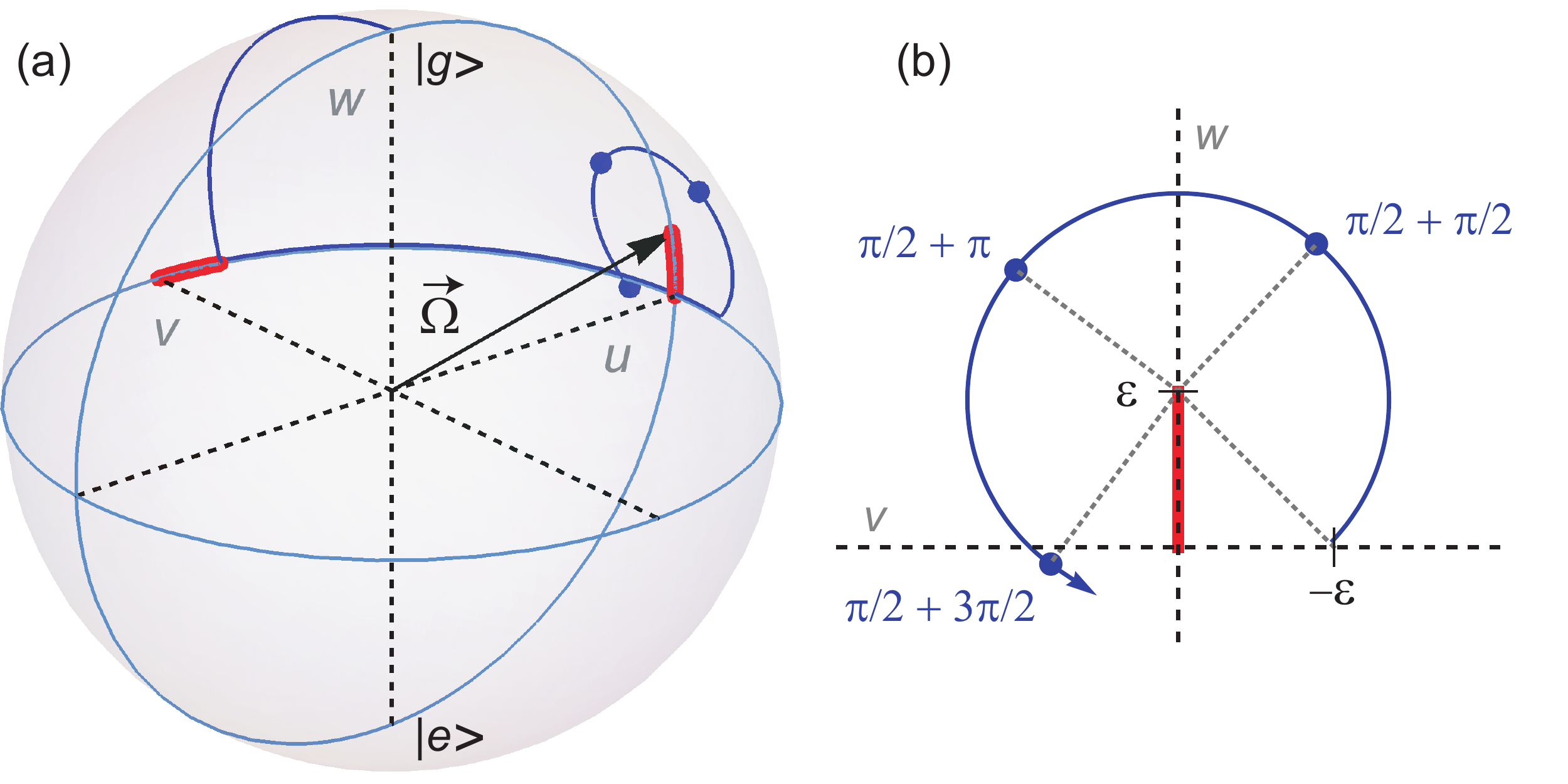}
\caption[]{Computed Bloch vector trajectory for a composite pulse sequence with a $\pi/2$ preparation pulse and an up to $3\pi/2$ long readout pulse assuming $\delta = 0$, $\varepsilon = 0.2$, and $\phi^{+} = \pi/2$. (a) The system is initialized in the ground state $|g\rangle$ and evolves under a sequence of coherent operations. (b) A magnified projection of the final part of the trajectory onto the $v$-$w$ plane shows how a $3\pi/2$ readout pulse suppresses the mapping of a $\delta'$-induced offset along the $v$-axis to a clock error-inducing offset along the $w$-axis. The $w$ coordinate is related to the excited state population $p$ via $w = 1-2p$. All spectroscopic feedback is solely determined by the measured populations, i.e., by the $w$-axis projection of the final Bloch vector.\label{f:traj3d2d}}
\end{center}
\end{figure}

To clarify the effects of the tilt- and speed-terms it is insightful
to visualize the system's unitary evolution on the Bloch sphere as shown
in Figure \ref{f:traj3d2d} for the $\phi^{+}$ case.
With the standard conversion \cite{Nielsen2000} from
the complex state vector $(g(t),e(t))$ to a real-valued three-dimensional Bloch vector $(u(t),v(t),w(t))$ the driven motion of the system corresponds to a rotation of the Bloch vector through an angle $\Omega_{0} \tau \sqrt{1 + \varepsilon^2}$ about the field vector $\mathbf{\Omega} = \Omega_{0} \hat{\mathbf{e}}_{u} + \delta' \hat{\mathbf{e}}_{w}$.
Therefore a $\delta'$-defect
will manifest itself after the initial $\pi/2$-pulse via the tilt-term to first order in an azimuthal
displacement of the Bloch vector on the equatorial plane by an angle
$-\varepsilon$ (horizontal red line segment). Displacements
due to the speed-factor are only of second order in $\varepsilon$.

After the subsequent $\phi^{+} = \pi/2$ phase hop (corresponding to a $-\pi/2$ in-plane rotation)
the Bloch vector should now ideally be oriented along $\hat{\mathbf{e}}_{u}$ but because of the initially acquired azimuthal offset it
is further inclined by an angle of $-\varepsilon$ to the median $u$-$w$ plane. From there it will resume its rotation
about $\mathbf{\Omega}$ as displayed in detail in Figure \ref{f:traj3d2d}(b). Since the field vector itself is tilted at an angle $\varepsilon$
out of
the equatorial plane (vertical red line segment) it is clear that a conventional
$\pi/2$ readout pulse will move the Bloch vector to a final position (marked by the first blue dot on the example trajectory) which is $2 \varepsilon$ away from the equator where $p^{+} = 1/2$. In a sequence with nonzero Ramsey time $T$ such a first-order displacement in $w$-direction (encoding the excited state population) would then erroneously be interpreted to indicate a frequency deviation $\omega_{\mathrm{LO}} \neq \omega_{eg}$ eventually leading to a $\delta'$-induced first-order clock error $\delta = -2 \varepsilon / T$. For a $3 \pi/2$ readout pulse the geometric situation is different: The Bloch vector continues to move another two quarter-turns on its small cone trajectory around $\mathbf{\Omega}$
so that it finally returns (ignoring higher-order speed term effects) to the equatorial plane.

Altogether, a total atom-light interaction time of $2 \pi / \Omega_{0}$ ensures that a
false drive frequency $\omega_{\mathrm{LOdrive}} \neq \omega'_{eg}$ does not lead to a first-order change in the measured excited state populations $p^{\pm}$. The population deviation latently acquired in the first half-time is
coherently compensated in the second half-time. Assuming that the false $\omega_{\mathrm{LOdrive}}$ is
unchanged for all pulse segments
this implicit common-mode suppression
corresponds to the explicit common-mode suppression realized in an incoherent way
with auto-balanced Ramsey spectroscopy. Note that
the population sensitivity $p^{\pm}(\delta)$ to frequency offsets $\delta$ during an inserted dark Ramsey interval
is not
coupled to the sensitivity of the population $p^{\pm}(\delta')$ to drive frequency deviations $\delta'$. If the Ramsey interval is added at a proper
point
during the $2 \pi$ long interrogation sequence (when the Bloch vector is crossing the equatorial plane) the frequency error signal $\tilde{p}(\delta)$ will exhibit the same steepness and contrast
as with the standard Ramsey protocol for equally long Ramsey times. Reference \cite{Falke2012} discusses
sensitivity aspects of various spectroscopy configurations.

\section{Universal immunity via incoherent balancing}

Coherent balancing as discussed above
suppresses the linear
error propagation from $\delta'$ to $\delta$. Since symmetry arguments \cite{Footnote01}
forbid second-order contributions to $\delta(\delta')$, one
expects a
cubic scaling of the induced clock error $\delta$
with the drive detuning $\delta'$ as described and observed in \cite{Yudin2010,Huntemann2012hrs}. Auto-balanced Ramsey spectroscopy
eliminates any residual dependance
of $\delta$ on $\delta'$, but more importantly, because of its nonspecific nature it simultaneously addresses all interrogation-induced clock shifts.

While some recently
proposed coherent Hyper-Ramsey
derivatives \cite{Hobson2016,Zanon2016} can
also completely
decouple $\delta$ from $\delta'$, they are
specifically tailored
to minimize light shift induced clock errors, do not preserve fringe symmetry,
and do not cover other interrogation defects. Reference \cite{Yudin2016} suggests the construction of "synthetic" clock frequencies by exploiting the functional dependance of $\delta$ on the Ramsey time $T$ for a given detuning $\delta'$. Depending on the number of
different Ramsey times these synthetic frequencies develop a corresponding higher order immunity.

Note
that all spectroscopy scenarios considered here are strictly speaking hybrid concepts that combine spectroscopic information not exclusively in a coherent or incoherent way, e.g., $\tilde{p}$ is always incoherently derived from $p^{+}$ and $p^{-}$. The
classification is merely based on the intrinsic balancing or shift suppression mechanism.

\section{Clock stability implications}

Whether the shift suppression is implemented
in a coherent or incoherent
way has distinct consequences for the stability of the atomic clock.
First consider a standard Ramsey clock operation with dark time $T$, Ramsey detuning $\delta$, and independent drive detuning $\delta'$. Even under ideal conditions, i.e., with constant $\delta' = 0$, zero mean detuning $\langle\delta\rangle = 0$, and negligible overhead time, the feedback-controlled LO frequency $\omega_{\mathrm{LO}}$ exhibits
residual frequency fluctuations $\Delta \omega_{\mathrm{LO}}$ caused by the finite signal-to-noise ratio of each population measurement.
Besides technical issues, quantum projection noise \cite{Itano1993} fundamentally limits the attainable signal-to-noise ratio. Given that projection measurement noise is unchanged for different dark times it is
convenient to introduce
a $T$-independent Ramsey phase noise $\Delta \phi_{\mathrm{LO}}$ and express the observed LO frequency fluctuations as $\Delta \omega_{\mathrm{LO}} = \Delta \phi_{\mathrm{LO}} / T$. In the context of an auto-balanced Ramsey interrogation this $T$-independence
indicates that the injected phase correction $\phi^{\mathrm{c}}$ (itself derived from an uncorrelated equally noisy short Ramsey sequence)
effectively doubles the frequency noise variance observed in the long Ramsey sequence. In other words,
due to phase noise on $\phi^{\mathrm{c}}$ carried over from the short into the long Ramsey sequence one expects the $\omega_{\mathrm{LO}}$-deviations in the auto-balanced long Ramsey sequence
to be a factor of $\sqrt{2}$ larger than those observed with an equally long injection-free sequence:
Combining two not coherently connected interrogation segments is stability-wise
inferior to merging the segments in a coherent
way so that projection noise is not encountered twice.

On the other hand, to minimize interrogation-induced shifts it is
advisable to only employ
low-power drive pulses, which implies longer pulse durations. With additional constraints (limited laser coherence time) for the total length of the interrogation sequence one
typically has to compromise between pulse duration and Ramsey time. For example, the Yb$^{+}$ E3 clock described in \cite{Huntemann2012hrs} relies on a Hyper-Ramsey interrogation scheme and is operated with a $\pi/2$-pulse duration to Ramsey time ratio of one-to-four. By replacing the $3\pi/2$ long composite readout pulse with a simple $\pi/2$-pulse the available Ramsey time $T$ is increased by
50 percent.
The longer $T$-interval reduces the clock instability by a corresponding factor of $3/2$, i.e.,
the $\sqrt{2}$ stability
drop due to auto-balancing is
outweighed by the gained
Ramsey time.

The above example
demonstrates
that using an
auto-balanced version of Ramsey spectroscopy can be advantageous
not only in terms of accuracy
but also in terms of clock stability.
In particular when considering non-ideal
scenarios (with significant initialization times, intrinsically unstable local oscillators, and additional technical constraints) the
stability performance of the various interrogation schemes will depend
on specific details of the
investigated system.


\end{document}